\documentclass[12pt]{iopart}
\usepackage{iopams}
\usepackage{epsf,graphics,graphicx}
\usepackage{float}
\usepackage{subfigure}
\newcommand{\ket}[1]{|#1\rangle}
\newcommand{\bra}[1]{\langle #1|}

\begin{document}
\title{Universal non-adiabatic holonomic gates in quantum dots and single-molecule 
magnets}
\author{Vahid Azimi Mousolou$^1$}
\ead{vahid.mousolou@lnu.se}
\author{Carlo M. Canali$^1$}
\ead{carlo.canali@lnu.se}
\author{Erik Sj\"oqvist$^{2,3}$}
\ead{erik.sjoqvist@kvac.uu.se}
\vskip 0.5 cm
\address{$^1$Department of Physics and Electrical Engineering, Linnaeus University,
391 82 Kalmar, Sweden}
\address{$^2$Department of Quantum Chemistry, Uppsala University, Box 518,
SE-751 20 Uppsala, Sweden}
\address{$^3$Centre for Quantum Technologies, National University of Singapore,
3 Science Drive 2, 117543 Singapore, Singapore}
\date{\today}
\begin{abstract}
Geometric manipulation of a quantum system offers a method for fast, universal, and 
robust quantum information processing. Here, we propose a scheme for universal 
all-geometric quantum computation using non-adiabatic quantum holonomies.  
We propose three different realizations of the scheme based on an unconventional 
use of quantum dot and single-molecule magnet devices, which offer promising 
scalability and robust efficiency.
\end{abstract}
\submitto{\NJP}
\maketitle
\section{Introduction}
Holonomic quantum computation (HQC), originally  conceived by Zanardi and Rasetti 
\cite{zanardi99}, has become one of the key approaches to perform robust quantum 
computation. The idea of HQC is based on the Wilczek-Zee holonomy \cite{wilczek84} 
that generalizes the Berry phase \cite{berry84} to non-Abelian (non-commuting) geometric 
phases accompanying adiabatic evolution. It was shown in Ref.~\cite{zanardi99} that 
adiabatic quantum holonomies generically allow for universal quantum computation. 
Different physical settings for adiabatic HQC have been proposed \cite{duan01,faoro03,solinas03}. 

Recently, universal HQC based on Anandan's non-adiabatic non-Abelian geometric phase 
\cite{anandan88} has been proposed \cite{sjoqvist12}. This scheme allows for high-speed 
implementations of quantum gates; a feature that can be particularly useful in solid-state 
devices, where qubits typically have short coherence time. Non-adiabatic HQC can be made 
robust to a wide class of errors, such as decay, dephasing, and fluctuations in control 
parameters \cite{johansson12}, and can be realized in decoherence free subspaces and 
noiseless subsystems insensitive to various collective errors \cite{xu12,zhang13}. First 
experimental demonstrations of the scheme for universal non-adiabatic HQC proposed 
in Ref. \cite{sjoqvist12} have been carried out for a superconducting qubit \cite{abdumalikov13} 
and in an NMR system \cite{feng13}. 

Here, we develop a setting  for conditional holonomic gates in a four-level 
configuration. We illustrate how non-adiabatic holonomic gates in our scheme can be 
implemented in solid state devices consisting of few-electron quantum dots \cite{kouwenhoven01} 
and single-molecule magnets (SSMs) \cite{gatteschi_sessoli}, which are promising candidates 
for quantum computation \cite{leuenberger01,tejada01,hanson07,zac10}. By utilizing the scalability of these 
systems, the geometric nature of the proposed holonomic gates is a feature that may allow 
for robust and fast manipulations of a large number of qubits.  

The outline of the paper is as follows. In the next section, we demonstrate one- and two-qubit 
quantum gates based on non-adiabatic quantum holonomies. The scheme is based on a 
generic four-level model with off-diagonal $2\times 2$ couplings. Realizations of the 
model in various kinds of solid-state devices are identified and  examined in Sec. 
\ref{sec:implementations}. The paper ends with the conclusions. 

\section{Non-adiabatic quantum holonomy}
The general structure of our system is described by a four-dimensional effective state 
space whose dynamics is governed by a Hamiltonian of the form 
\begin{eqnarray}
H(t) = \hbar \Omega (t) \left( \begin{array}{rr}
0 & T \\
T^{\dagger} & 0 
\end{array} \right) ,
% \  T=\left( \begin{array}{rr}
%T_{11} & T_{12}\\
%T_{21} & T_{22} \\
%\end{array} \right)  
\label{eq:hamiltonian}
\end{eqnarray}
in the ordered orthonormal basis $\mathcal{M}=\{\ket{a}, \ket{b}, \ket{c}, \ket{d}\}$. 
We assume that $T$ is a complex-valued and time-independent $2 \times 2$ matrix 
that satisfies $\det T \neq 0$, where the latter  ensures that $T$ is invertible. 
The Hamiltonian is turned on and off as described by the time-dependent scaling function 
$\Omega (t)$. We demonstrate in the following how $H(t)$ can be used for universal 
non-adiabatic HQC and how it can be realized in systems of four coupled quantum dots 
or in triangular molecular antiferromagnets. 

The Hamiltonian in Eq. (\ref{eq:hamiltonian})  naturally splits the state space into 
two subspaces spanned by $\mathcal{M}_0 = 
\{\ket{a},\ket{b} \}$ and $\mathcal{M}_1=\{\ket{c},\ket{d} \}$. Non-adiabatic holonomy 
transformations are realized if (i) the Hamiltonian vanishes on each evolving subspace, and  
(ii) the projection operators $P_0 = \ket{a} \bra{a} + \ket{b} \bra{b}$ and $P_1 = \ket{c} 
\bra{c} + \ket{d} \bra{d}$ move in a cyclic fashion. Condition (i) is satisfied since  
$P_l (t) H(t) P_l (t)=U(t,0)P_l H(t) P_l U^{\dagger} (t,0) =0$, $l=0,1$, where the 
first equality follows from the fact that $H(t)$ commutes with the time evolution 
operator $U(t,0)= e^{-(i/\hbar) \int_0^t H(s) ds}$. For condition (ii), we first note that since $T$ is 
assumed to be invertible, there is a unique singular value decomposition $T = U_0 D U_1^{\dagger}$, 
where $D = {\textrm{diag}} \{ \alpha , \beta \}$ and $U_0,U_1$ are the positive and unitary 
parts, respectively. By using the singular value decomposition, the time evolution 
operator can be expressed as 
\begin{eqnarray}
U(t,0) & = & \left( \begin{array}{cc} 
U_0 \cos \left( a_t D  \right) U_0^{\dagger} & -iU_0 \sin \left( a_t D  \right) U_1^{\dagger}\\ 
-iU_1 \sin \left( a_t D \right) U_0^{\dagger}  & U_1 \cos \left( a_t D \right) U_1^{\dagger} 
\end{array} \right)\hspace{0.5cm}
\label{eq:u(t)}
\end{eqnarray}
where $a_t = \int_0^t \Omega (s) ds$. The evolution of the $P_l$'s is thus cyclic provided 
there is a $t=\tau$ such that $\sin \left( a_{\tau} D \right) = {\textrm{diag}} \{ \sin (a_{\tau} \alpha), 
\sin (a_{\tau} \beta) \} = 0$. Under this condition the 
holonomy matrices for the two subspaces 
$l=0,1$ read
\begin{eqnarray}
U(\mathcal{C}_ {l}) & = & U_l Z^n U_l^{\dagger} , \ n=0,1,
\label{eq:holonomy} 
\end{eqnarray}
where $\cos \left( a_{\tau} D  \right)=Z^n$ with $Z ={\textrm{diag}} \{1,-1\}$. $\mathcal{C}_0$ 
and $\mathcal{C}_1$ are the paths of the two subspaces spanned by $\mathcal{M}_0$ and 
$\mathcal{M}_1$, respectively. These paths reside in the Grassmannian manifold 
$G(4;2)$, being the space of all two-dimensional subspaces of the four-dimensional state space. 
One can see from Eq. (\ref{eq:holonomy}) that the holonomy is non-trivial if and only if $n=1$. 

Next, we encode the basis states in $\mathcal{M}$ into the computational two-qubit states 
$\ket{a} \rightarrow \ket{00}$, $\ket{b} \rightarrow \ket{01}$, $\ket{c} \rightarrow \ket{10}$, 
and $\ket{d} \rightarrow \ket{11}$. Running the Hamiltonian $H(t)$ along certain time intervals 
for appropriate choices of $T$ provides a realization of holonomic conditional gates of the form  
\begin{eqnarray}
U(\mathcal{C}_0,\mathcal{C}_1) = \ket{0} \bra{0} \otimes U(\mathcal{C}_0) + 
\ket{1} \bra{1} \otimes U(\mathcal{C}_1) ,  
\end{eqnarray}
where the holonomies $U(\mathcal{C}_0)$ and $U(\mathcal{C}_1)$ act on the target qubit
conditionalized on the states $\ket{0}$ and $\ket{1}$, respectively, of the control qubit. One 
can see that by switching the roles of control and target qubits, which can be done by swapping 
the encoding of states $\ket{b}$ and $\ket{c}$, holonomic conditional two-qubit gate of the 
form $U(\mathcal{C}_0)\otimes\ket{0}\bra{0} + U(\mathcal{C}_1)\otimes\ket{1}\bra{1}$ can 
be carried out. 

We now verify that $U(\mathcal{C}_0,\mathcal{C}_1)$ is sufficient for implementing a universal 
set of one- and two-qubit gates. The key observation is that any triplet $U_0,U_1,$ and $D$ 
can be realized since $T$ is arbitrary up to the condition $\det T \neq 0$. First, by choosing 
$T > 0$ such that $\cos  \left( a_{\tau} D \right) = Z$, it follows that $U_1 = U_0$ and 
$U(\mathcal{C}_0,\mathcal{C}_1)$ reduces to the one-qubit holonomic gate 
$U_0 Z U_0^{\dagger} = {\bf n} \cdot {\bf R}$ with ${\bf R} = (X,Y,Z)$ the Pauli 
operators acting on the target qubit and ${\bf n}$ a unit vector. Any given ${\bf n}$, 
characterized by the spherical polar angles $\theta,\phi$, 
can for instance be realized by choosing $T=U_0DU_0^{\dagger}$ and $a_{\tau}$, such 
that $a_{\tau}D = \pi {\textrm{diag}} \{ 2,1 \}$ and $U_0=e^{-i\phi Z/2} e^{-i\theta Y/2}$. 
By applying sequentially two such gates with different $T$'s corresponding to ${\bf n}$ 
and ${\bf m}$ yields $({\bf m} \cdot {\bf R})({\bf n} \cdot {\bf R}) = {\bf n} \cdot {\bf m} - 
i ({\bf n} \times {\bf m}) \cdot {\bf R}$, which is an arbitrary SU(2) transformation acting 
on the target qubit. Thus, our setup allows for arbitrary holonomic one-qubit gates.  
Secondly, by choosing $T$ with $U_1 \neq U_0$ and $\cos  \left( a_{\tau} D \right) = Z$ 
implements the holonomic conditional gate $U(\mathcal{C}_0,\mathcal{C}_1) = \ket{0} \bra{0} 
\otimes U_0 Z U_0^{\dagger} + \ket{1} \bra{1} \otimes U_1 Z U_1^{\dagger}$ that may entangle 
the qubits. Any entangling two-qubit gate is universal when assisted by arbitrary 
single-qubit transformations \cite{brenner02}. In other words, our scheme is universal. 

Conceptually, the proposed holonomic conditional gate $U(\mathcal{C}_0,\mathcal{C}_1)$ 
can be viewed as a non-Abelian version of the Abelian geometric phase shift gate developed 
in Ref. \cite{wang01,zhu02,zhu03}. In the Abelian case, evolution is chosen such that two orthogonal 
qubit states $\ket{0}$ and $\ket{1}$, say, pick up zero dynamical phases (for instance, by 
parallel transporting $\ket{0}$ and $\ket{1}$), which implies that $\ket{0}\rightarrow 
e^{i\gamma_g} \ket{0}$ and $\ket{1}\rightarrow e^{-i\gamma_g} \ket{1}$, $\gamma_g = 
\frac{1}{2} \Omega$ being the Aharonov-Anandan (AA) geometric phase \cite{aharonov87} 
with $\Omega$ the solid angle enclosed on the Bloch sphere.  This defines the AA geometric 
phase gate $\ket{k} \rightarrow e^{i(1-2k) \gamma_g} \ket{k}$, $k=0,1$. In our non-Abelian 
scheme, conditional two-qubit gate is implemented if there exist pair of loops $\mathcal{C}_0$ 
and $\mathcal{C}_1$ in the Grassmannian manifold $G(4;2)$ of the target qubit along which 
there is no dynamical phase. Note that the orthogonal subspaces $\mathcal{M}_0$ and 
$\mathcal{M}_1$ correspond to the state space of the target qubit when the control qubit 
is in the state $\ket{0}$ or $\ket{1}$, respectively. Zero dynamical contribution to the evolution 
of these two orthonormal subspaces is a signature of the geometric nature of our holonomic 
conditional two-qubit gate, just as zero dynamical phase in the Abelian schemes 
\cite{wang01,zhu02,zhu03} is a signature of the geometric nature of the above AA gates. 

\section{Physical implementations}
\label{sec:implementations}
\subsection{Coupled quantum dots}
\subsubsection{Tight-binding model}
As possible physical implementations of the holonomic gates, we here discuss two settings 
involving four coupled quantum dots arranged in closed ring geometry. The first setting  
involves the well-known two-dimensional electron gas supported, e.g., by  a GaAs/AlGaAs 
heterostructure, depleted by a convenient set of top gate electrodes to form four quantum 
dots where electrons are confined. Devices of this type can be routinely fabricated nowadays 
with state of the art electron beam lithography \cite{petersson10}. In this system, electrons 
can tunnel from one dot to another with a tunnel coupling controlled individually by inter-dot 
gates. We also assume that the local potential, and hence the available single-particle energy 
levels in each quantum dot, can be individually manipulated with back gates. HQC is realized 
when this four-quantum-dot device contains just a single electron. 
If only one single-particle orbital in each quantum dot is relevant,
the system can be described by a four-site tight-binding model, forming a closed ring. The 
second-quantized Hamiltonian reads
\begin{eqnarray}
H = \sum_{k=1,\sigma}^4 \epsilon_{k}c_{k,\sigma}^{\dagger} c_{k,\sigma}^{\phantom\dagger} + 
\sum_{k=1}^4 \left[t_{k,k+1} c_{k,\sigma} ^{\dagger}c_{k+1,\sigma}^{\phantom\dagger}+h.c.\right],\ \ 
\label{eq:hamiltonian1}
\end{eqnarray}
where $c_{k\sigma}^{\dagger}$ ($c_{k\sigma}^{\phantom\dagger}$) are electron creation (annihilation) operators 
for Wannier orbitals localized at dot-site $k$, with on-site energy $ \epsilon_{k}$ 
and spin quantum number $\sigma = \uparrow, \downarrow$; 
we assume periodic boundary conditions
$c_{5,\sigma}=c_{1,\sigma}$. 
The $t_{k,k+1}$ are nearest-neighbor 
hopping parameters, which can be controlled by gate voltages. 

In the presence of an external magnetic field $B$, the Hamiltonian in
Eq.~(\ref{eq:hamiltonian1}) acquires the Zeeman term
$-g B\sum_{k\sigma}\sigma c_{k\sigma}^{\dagger} c_{k\sigma}^{\phantom\dagger}$,
where the Bohr magneton has been absorbed into  $g$. Furthermore, the magnetic field 
generates a gauge-invariant modification of the tight-binding hopping parameters
via the Peierls phase factors \cite{graf95,ismail01,cehovin04}
\begin{eqnarray}
t_{k,k+1} \rightarrow  J_{k,k+1}=t_{k,k+1}e^{-i\frac{e}{\hbar} {\alpha_{k,k+1}}}\; ,
\label{eq:hopping}
\end{eqnarray}
where $\alpha_{k,k+1}=\int_{k\rightarrow k+1} \mathbf{A} \cdot d \mathbf{l}$ 
is the integral of the
vector potential along the hopping path from site $k$ to site $k+1$.

We fix the on-site energies $\epsilon_{k}=0 $ by adjusting the voltage on the back gates 
of the device. We now assume that  hopping parameters can be turned on and off with the 
same time-dependent scaling function $\Omega (t)$ by sweeping the top gates. In the 
invariant Fock subspace spanned by the four (spin-polarized) one-electron states
$\{\ket{\uparrow\!000},\ket{00\!\uparrow\!0},\ket{0\!\uparrow\!00},\ket{000\!\uparrow}\}$, 
the time dependent Hamiltonian in this ordered basis has the form of 
Eq.~(\ref{eq:hamiltonian}), with the time-independent $2\times2$ submatrix
\begin{eqnarray}
T = \left( \begin{array}{cc}
\mathcal{J}_{1,2} & \mathcal{J}_{4,1}^{\ast} \\
\mathcal{J}_{2,3}^{\ast} & \mathcal{J}_{3,4} \\
\end{array} \right) ,
\label{p-hamiltonian}
\end{eqnarray}
where we have put $J_{k,k+1} = \hbar \Omega (t) \mathcal{J}_{k,k+1}$. Note that for a 
given magnetic field we must choose the complex-valued hopping parameters so that 
their phases $\alpha_{k,k+1}$ add up to the  total magnetic flux $\Phi = \oint \mathbf{A} 
\cdot d \mathbf{l}= {\cal A} B$, where $\cal A$ is the area enclosed by the four-dot ring. 
A change of these phases corresponds to a gauge transformation $\mathbf{A} \rightarrow 
\mathbf{A} + \nabla\Lambda$ under which the holonomies of the two subspaces $l=0,1$ 
transform into $U(\mathcal{C}_ {l})\rightarrow\Gamma_ {l} U(\mathcal{C}_ {l}) 
\Gamma_ {l}^{\dagger}$, where $\Gamma_ {l} = {\textrm{diag}} \{e^{-i\frac{e}{\hbar} 
\Lambda(l+1)}, e^{-i\frac{e}{\hbar}\Lambda(l+3)}\}$ with $\Lambda(k)$ indicating the 
value of $\Lambda$ at site $k$. This shows the holonomies associated with this 
field-dependent tight-binding Hamiltonian are gauge covariant quantities. 
 
In order to take advantage of non-adiabaticity to control decoherence 
effects, the typical switching time $\tau$ should be as short as possible. At the same 
time, the action of the Hamiltonian must be "adiabatic" enough to prevent transitions to higher 
orbital levels not included in Eq.~(\ref{eq:hamiltonian1}). This requirement is expressed by 
the condition $\tau \gg \hbar/ \Delta \epsilon$, where $\Delta \epsilon$ is the 
single-particle level spacing. For a quantum dot with a size of 100 nm, typically 
$\Delta \epsilon\sim 1 {\rm meV}$, which would permit $\tau$ 
to be as fast as 10 ps. This time is considerably shorter than the phase coherence time
for charge qubits in quantum dots ($\sim$ 1-10 ns \cite{petersson10}).

\subsubsection{Spin model}
As the second setting to realize our scheme, we start again with the four-dot system 
introduced above, but we consider the case where four electrons are confined in the device. 
Now,  electron-electron interactions must be included and we can do so by adding to the 
Hamiltonian of Eq.~(\ref{eq:hamiltonian1}) an onsite Hubbard repulsion term
$U\sum_k n_{k\uparrow} n_{k \downarrow}$, where 
$n_{k \sigma} = c_{k\sigma}^{\dagger} c_{k\sigma}^{\phantom\dagger}$. In the limit 
$t_{k,k+1}/U \ll 1$ double occupancy in each dot is suppressed and the low-energy physics 
of the system can be described by the spin Hamiltonian \cite{trif10}
\begin{eqnarray}
\tilde{H}_{\textrm{s}} & = &\sum_{k=1}^4 
\tilde{J}_{k,k+1} {\bf s}_{k}\cdot{\bf s}_{k+1}  + 
\sum_{k=1}^4{\bf s}_{k}\cdot \Theta_{k,k+1}{\bf s}_{k+1} 
\nonumber\\
 & & + \sum_{k=1}^4\tilde{{\bf D}}_{k,k+1}\cdot\left({\bf s}_k \times {\bf s}_{k+1}\right) , 
\label{hamiltonian2}
\end{eqnarray}
where ${\bf s}_k$ is a spin$-\frac{1}{2}$ vector operator localized at site $k$. In 
Eq. (\ref{hamiltonian2}), the first term is an isotropic Heisenberg interaction, with 
antiferromagnetic exchange coupling $\tilde{J}_{k,k+1}$, the second term is an anisotropic 
exchange described by second rank symmetric tensors $\Theta_{k,k+1}$, and the last term 
is an antisymmetric Dzialoshinski-Moriya (DM) interaction. 
The general exchange coupling arises from the interplay between the 
electron-electron repulsion and spin-orbit interaction. 
For the geometry of four coplanar quantum dots in the 
$xy$ plane only $z$ components of the DM exchange vectors $\tilde{{\bf D}}_{k,k+1}$ 
are relevant. In an in-plane electric field, the parameters in Eq. (\ref{hamiltonian2}) can be modified so that the effective Hamiltonian of the spin-model is given by 
\begin{eqnarray}
H_{\textrm{s}} =  \sum_{k=1}^4 
{\bf s}_{k}\cdot\bar{\bar{J}}_{k,k+1}{\bf s}_{k+1} 
+\sum_{k=1}^4D_{k,k+1}^{z}\cdot\left({\bf s}_k \times {\bf s}_{k+1}\right)_z.\hspace{5mm} 
\label{spin-hamiltonian2}
\end{eqnarray}
Here, we have $\bar{\bar{J}}_{k,k+1}=\textrm{diag}\{J_{k,k+1}, J_{k,k+1}, 0\}$ with  $J_{k,k+1} \sim t_{k,k+1}^2/U$ and ${D}^z_{k,k+1}\sim t_{k,k+1}\lambda_{k,k+1}/U$, where $\lambda_{k,k+1}$ is a spin-dependent hopping term describing the spin-orbit interaction.

Let $\ket{\uparrow}_k$ $(\ket{\downarrow}_k)$ be the eigenvector of the 
$z$ component of the spin operator at site $k$ corresponding to the eigenvalue 
$\frac{1}{2}$ $(-\frac{1}{2})$. The four-dimensional subspace with $S_{\rm total}^z =1$, 
spanned by the ordered basis
$\mathcal{M}=\{\ket{\downarrow\uparrow\uparrow 
\uparrow},\ket{\uparrow\uparrow\downarrow\uparrow},\ket{\uparrow\downarrow 
\uparrow\uparrow},\ket{\uparrow\uparrow\uparrow\downarrow} \}$ is an invariant subspace
\cite{note1}. 
To realize the HQC scheme, the parameters  $J_{k,k+1}$ and $D_{k,k+1}^{z}$ 
have to be turned on and off with the same time-dependent function $\Omega (t)$.
Under this condition, the Hamiltonian in the ordered basis
$\mathcal{M}$ has exactly the form of 
Eq.~(\ref{eq:hamiltonian}) with the time-independent submatrix 
\begin{eqnarray}
T = \left( \begin{array}{cc} 
\mathcal{J}_{1,2}-i\mathcal{D}_{1,2}^{z}& \mathcal{J}_{4,1}+i\mathcal{D}_{4,1}^{z}\\
\mathcal{J}_{2,3}+i\mathcal{D}_{2,3}^{z}&\mathcal{J}_{3,4}-i\mathcal{D}_{3,4}^{z} \\
\end{array} \right),
\label{T_submatrix}
\end{eqnarray}
where we have put $J_{k,k+1} = \hbar \Omega (t) \mathcal{J}_{k,k+1}$ and $D^z_{k,k+1} = 
\hbar \Omega (t) \mathcal{D}^z_{k,k+1}$. It is important to stress that since both $J_{k,k+1}$ 
and $D^z_{k,k+1}$ are proportional to the 
inter-dot tunneling constants, they can be efficiently manipulated with time-dependent gate 
voltages. 

\subsection{Single-molecule magnet}
The spin  Hamiltonian in Eq.~(\ref{hamiltonian2}) also describes 
the magnetic properties of a class of single-molecule magnets (SSMs) composed of
antiferromagnetic spin rings \cite{gatteschi_sessoli}. Such systems would in principle be more 
scalable, compact, and reproducible than the devices based on quantum dots, since their 
exchange constants and spin coherence properties are controlled by the molecular structure 
of the molecule, which can be chemically engineered \cite{ardavan_2012}.

The systems that we are interested in are triangular 
SMMs such as $\textrm{Cu}_3, \textrm{V}_{15}, \textrm{Co}_3$ \cite{kortz01,choi06}.
Their magnetic core consists of three $s= 1/2$ spins positioned at the vertexes 
of an equilateral triangle
and coupled by an antiferromagnetic Heisenberg exchange interaction. The ground state (GS) manifold
of these frustrated SMMs
is given by two degenerate {\it total spin} $S=1/2$ doublets ($M = \pm 1/2$) of opposite 
spin chirality $\chi = \pm 1$. 
The energy
gap $\Delta_J$ to the first excited $S= 3/2$ quadruplet is typically in the order of 1 meV. 
A spin-orbit induced DM interaction lifts the degeneracy between the two chiral doublets with
a splitting $\Delta_{\textrm{SO}} \le 0.1  \Delta_J$ \cite{trif08, Nossa12}. 
What makes these triangular SMMs interesting for quantum manipulation is that, 
due to the lack of inversion symmetry, an electric field in the $xy$-plane of the molecule couples the
two GS doublets of opposite chirality \cite{trif08, trif10, Islam10}. 
In the four-dimensional GS manifold spanned by the
states $|\chi= \pm 1, M = \pm 1/2\rangle$, the effective low-energy spin Hamiltonian in the
presence of external electric $\bf E$ and magnetic $\bf B$ fields is \cite{trif08}
\begin{eqnarray}
H_{\textrm{eff}}=\Delta_{\textrm{SO}}C_zS_z+
d\;\textbf{E}\cdot\textbf{C}_{\parallel}+
\textbf{B}\cdot\bar{\bar{g}}\textbf{S},
\label{eff. hamiltonian}
\end{eqnarray}
where $\textbf{C}=(\{\textbf{C}_{\parallel}=C_x, C_y\},  C_z)$ and $\textbf{S}=(S_x, S_y, S_z)$ 
are respectively the chirality and spin operators. Due to the symmetry of the molecule 
$\bar{\bar{g}}=\textrm{diag}\{g_{\parallel}, g_{\parallel}, g_{\bot}\}$. The parameter $d$ is the 
effective electric dipole coupling, and it gives the strength of the coupling between the two
states with opposite chirality brought about by the electric field. 
In Cu$_3$ SMMs $d$ is not small \cite{Islam10},
and for typical electric fields generated by scanning tunneling microscope ($\approx 10^2$ kV/cm) the
spin-chirality manipulation (Rabi) time is $10-10^3$ ps \cite{trif08, Islam10}.

\begin{figure}[h]
\centering
\includegraphics[width=70mm,height=45mm]{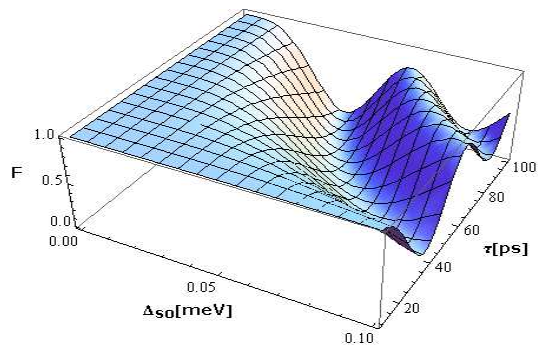}
\caption{Fidelity $F$ of the holonomic Y-gate for a triangular SMM
as a function of the spin-orbit induced splitting $\Delta_{\textrm{SO}}$ and operation time $\tau$. 
Here we have chosen 
$g_{\parallel} (B_{x} - iB_{y}) = 3 \hbar 
\Omega(t)$, $d(E_{x}-iE_{y})=-\frac{i}{2}\hbar \Omega(t)$, where $\Omega (t)$ is a square 
pulse such that $\int_0^{\tau} \Omega (t) dt=\pi$.}
\label{fig:fidelity}
\end{figure}

To implement our holonomic scheme in this setting, we assume ${\bf B}$ and ${\bf E}$ are 
turned on and off by a common scaling function $\Omega (t)$ and that $B_z$ is negligibly 
small. Under these conditions, the Hamiltonian in Eq. (\ref{eff. hamiltonian}) 
takes the form 
\begin{eqnarray}
H_{\textrm{eff}} = \hbar \Omega (t) \left( \begin{array}{cc}
0& T\\
T^{\dagger}&0\\
\end{array} \right) +\frac{\Delta_{\textrm{SO}}}{2}\left( \begin{array}{cc}
I& 0\\
0&-I\\
\end{array} \right) 
\label{eff. hamiltonian1}
\end{eqnarray}
in the ordered basis $\{\ket{+1,+1/2}$, $\ket{-1,-1/2}$, $\ket{+1,-1/2}$, 
$\ket{-1,+1/2}\}$ of the ground state subspace. Here, $T_{11}=T_{22}^{*} = 
\frac{1}{2}(\mathcal{B}_{x}-i\mathcal{B}_{y})$ and $T_{12}=T_{21}^{*}=(\mathcal{E}_{x} - 
i\mathcal{E}_{y})$, where we have put $g_{\parallel} B_k = \hbar \Omega (t) \mathcal{B}_k$ and 
$dE_k = \hbar \Omega (t) \mathcal{E}_k$, $k=x,y$.
This expression of the Hamiltonian shows that triangular SMMs such as Cu$_3$ can be 
used to implement efficient and robust gates. Although the restricted form of $T$ associated 
with this system does not allow for universal computation, we would 
still be able to obtain some of the most important gates such as all the Pauli-matrices and 
the controlled-Z gate in a fashion that is highly resistant to errors. The crucial challenge 
relies on the ability of generating electric and magnetic fields characterized by time-dependence 
function $\Omega(t)$. Fig.~\ref{fig:fidelity} shows the fidelity $F$ of the Pauli-Y gate performed 
with a triangular SMM, as a function of the spin-orbit induced splitting $\Delta_{\textrm{SO}}$ 
and operation time $\tau$. Here, we have adopted the definition \cite{note2}  
$F = F(W,V)=\frac{1}{2}+\frac{1}{8}{\textrm{Re}} \Tr \left( W^{\dagger}V \right)$,
where $W$ is the ideal gate and $V$ is the gate carried out by the perturbative 
Hamiltonian in Eq.~ (\ref{eff. hamiltonian1}).  
The fidelity $F(W,V)$ determines the 
Hilbert-Schmidt distance between $W$ and $V$. 
For typical values of $\Delta_{\textrm{SO}} \approx 0.02$ meV \cite{Nossa12} the fidelity is
over $98\%$ for $\tau=10-100$ ps.

It is worth mentioning that the spin model described by the Hamiltonian in Eq. (\ref{spin-hamiltonian2}) 
can be realized in other settings, such as magnetic impurities on the surface of topological insulators \cite{zhu11}, cold atoms in a two-dimensional optical lattice \cite{radic12}, and nuclear spin 
qubits interacting via a semiconductor-heterojunction \cite{mozyrsky01}. These systems may 
provide alternative platforms to implementing our holonomic gates.

\subsection{Scalability}
The four-level configuration presented above can be used in a scalable pattern necessary 
for universal quantum computation on any number of qubits. This can be achieved by 
implementing the Hamiltonian 
\begin{eqnarray}
\widetilde{H}(t) & = & 
\sum_{p<q} \hbar \Omega_{pq} (t) \left( \begin{array}{rr}
0\ \ \   & T^{(pq)} \\
T^{(pq)\dagger} & 0\ \ \ 
\end{array} \right) , 
\label{scalabe hamiltoinan}
\end{eqnarray} 
where $p$ and $q$ label the qubits. Any gate acting on qubits $p$ and $q$ can 
be implemented by turning on and off $\Omega_{pq} (t)$ while the other terms are kept off. 
$\widetilde{H}(t)$ can be realized with the first two examples using the techniques introduced 
in Ref. \cite{trifunovic12} to design scalable architectures in quantum dot systems. This 
Hamiltonian would also be realized by arranging triangular SMMs on a surface in a way that 
each of them can be manipulated individually. 

\section{Conclusions}
In conclusion, we have introduced a scheme to implement high-speed universal holonomic 
quantum gates that considerably differs from the other setups proposed for HQC. It allows 
for implementation of conditional gates using non-adiabatic quantum holonomies 
\cite{anandan88} acting on qubit pairs encoded in a four level structure. We make use of the 
two-qubit system as a unit cell for scalable all-geometric quantum computation. We propose 
three different realizations of our scheme based on an unconventional use of quantum dot 
and single-molecule magnet devices. We emphasize that the physical realizations of HQC 
presented here differ substantially from previous proposals of implementation of charge and 
spin qubits in quantum dots and single-molecule magnets. Finally, we would like to point 
out that the four level configuration proposed here is quite a general scheme that can be 
realized with other physical systems actively considered for quantum computer implementation
\cite{zhu11,radic12,mozyrsky01}. 
\section*{Acknowledgments}
This work was supported by School of Computer Science, Physics and Mathematics at Linnaeus 
University (Sweden). E.S. thanks National Research Foundation and the Ministry of 
Education (Singapore) for support.


\begin{thebibliography}{999}
\bibitem{note1} Note that this subspace involves the excited states with 
$S_{\rm total} = 1, 2$ of the spin Hamiltonian.
\end{thebibliography}
\end{document}